# A geometric method for spatiotemporal coherent structure analysis

Chuan Zhang[1,2], Kentaroh Takagaki[1,3], Xiaoying Huang[1], Steven J. Schiff[4] and Jian-Young Wu[1,*]

[1]*Department of Physiology and Biophysics, Georgetown University Medical Center, Washington, D.C. 20057, USA*
[2]*Department of Mathematics, Colorado State University, Fort Collins, Colorado 80523, USA*
[3]*Forschergruppe Neuroprothesen, Leibniz Institute for Neurobiology, Brenneckstrabe 6, Magdeburg 39118, Germany*
[4]*Center for Neural Engineering, Departments of Neurosurgery, Engineering Science and Mechanics, and Physics, The Pennsylvania State University, 212 Earth-Engineering Sciences Building, University Park, Pennsylvania 16802, USA*

**Abstract** We describe a geometric method to quantify wave patterns observed in the nervous system, which are non-stationary and with a mixture of spiral, target, plane and irregular waves. The method analyzes fluctuations of the energy angular distribution in two-dimensional Fourier spectrum of wave patterns, which reflects changes of the orientation distribution of wavefronts. We show that the number of the genuine peaks in generalized phase spectrum is close to the number of the coherent space-time clusters arising in wave patterns, and propose to use the number as a complexity measure.

## I. Introduction

Wave patterns with a mixture of spirals, plane, target and irregular patterns have been observed in both the brain and heart [1-4]. Quantifying these patterns is important and remains challenging. In the past few decades, several methods or measures, such as two-point correlation functions, extensive scaling of Lyapunov exponents and fractal dimensions etc., have been used to characterize spatiotemporal patterns [5]. Recently, the empirical orthogonal function (EOF) was used to characterize spatiotemporal patterns in neural tissue [6, 7]. The minimal number of eigenmodes required to account for the essential dynamical features of a spatiotemporal pattern was suggested to be used as a dimension measure [8]. While the EOF method is powerful in extracting coherent structures [6-10], it requires the extracted eigenmodes to be orthogonal to each other. This constraint is rarely reflected in the physical or biological patterns being analyzed, hence imposes a limit on the physical interpretability of each extracted eigenmode [11]. By using a binary reduction technique, P. Jung et. al. [14] decomposed spatiotemporal patterns into contiguous space-time regions, which were defined as coherent clusters, and showed the power-law cluster-size distribution for spiral turbulence.

In this paper, we investigate wave patterns recorded from neural tissues with the voltage-sensitive dye imaging (VSDI) technique [4], and extend the definition of the coherent space-time cluster to the space-time region(s) within which wave patterns have identical spatiotemporal symmetry (see **Appendix A**). Thus, one cluster may contain multiple spatially isolated regions. But due to the existence of non-local connectivity in neural tissues, the spatially isolated regions with identical spatiotemporal symmetry in the 2D plane may be just different "cross sections" of the same coherent cluster. Therefore, in this paper, we assume that two spatially isolated regions in the 2D plane are in the same coherent cluster, if wave patterns in the two regions have exactly the same spatiotemporal symmetry.

In actual data sets, boundaries of the clusters are highly dynamic and unclear, which causes technical difficulty in accurately identifying these cluster boundaries. To circumvent this difficulty, we inspect the characteristics of spatiotemporal symmetry of wave patterns, instead of identifying the clusters directly. Inspired by the spectral energy distribution properties of clusters with different spatiotemporal symmetries, we develop a geometric method to characterize the dynamics of wave patterns. We show that the number of genuine peaks (NGPs) provides an estimate of the number of clusters. By applying the method to both artificially generated and experimentally recorded wave patterns, we show that the method provides a more physically relevant measure for patterns typical of active neuronal systems.

## II. GP Spectrum Analysis

We investigate the spatiotemporal evolution of wave mixtures with rotational and non-rotational patterns. A coherent space-time cluster (for short, coherent cluster) is defined as space-time regions within which wave patterns have identical spatiotemporal symmetry (see **Appendix A**). In terms of the spatiotemporal symmetry of the wave patterns, coherent clusters can be categorized into two categories: translation invariant clusters and rotation invariant clusters. Wave patterns in the translation invariant clusters consist of traveling plane waves and emitting (absorbing) target waves, whereas wave patterns in the rotation invariant clusters comprise rotating spiral waves. A prominent characteristic that distinguishes these two categories is whether wave fronts change orientation. In a translation invariant cluster, the orientations of wave fronts do not change, whereas in a rotation invariant cluster the orientations of wave fronts change continuously.

Theoretically, any two-dimensional (2D) image can be decomposed into sinusoidal components in 2D space. Each sinusoidal component in physical space corresponds to a 2D Dirac delta function in wave number space, which lies in the wave vector direction of the sinusoidal component. Thus, the total integrated differential energy distributed in the same direction in wave number space corresponds to those of the sinusoidal components with the same corresponding orientation in physical space. According to Parseval's theorem, as long as the energy of the image components does not change, the corresponding spectral energy is also invariant. Therefore, the spectral energy of a translation invariant cluster keeps constant in all directions in wave number space due to the fixed orientations of its wave fronts. For a rotation invariant cluster, however, the spectral energy oscillates with identical frequency in all directions. If multiple clusters with different spatiotemporal symmetries coexist within the area being observed, due to the highly nonlinear nature of wave patterns observed in neuronal systems, origination, collision, and merging of wave fronts will occur frequently. These events cause changes in the spectral energy and its oscillation patterns. So intuitively, the fluctuation characteristics of the spectral energy of wave patterns shed light on its spatiotemporal complexity. The more organized the wave patterns are, the simpler the fluctuation of spectral energy will be. Conversely, the more disordered the wave patterns are, the more complex the fluctuation of spectral energy will be. Based on this intuition, we develop the following geometric method to quantify the complexity of wave patterns.

The amplitude of wave patterns is denoted by a scalar field $f(x,y,t)$, and its wave number spectrum is given by:

$$F(k_x,k_y,t) = \frac{1}{2\pi} \int_{-\infty}^{\infty} \int_{-\infty}^{\infty} f(x,y,t)\exp(-j(k_x x + k_y y))dxdy \qquad (1)$$

where $k_x$ and $k_y$ are wave numbers in $x$ and $y$ directions, respectively. Accordingly, the differential energy $dE$ is given by:

$$dE(k_x,k_y,t) = F(k_x,k_y,t) \cdot F^*(k_x,k_y,t) \qquad (2)$$

where $F^*(k_x,k_y,t)$ is the complex conjugate of $F(k_x,k_y,t)$. By integrating differential spectral energies distributed in same direction, the energy angular distribution (EAD) is obtained as:

$$\Phi(\theta,t) = \int_0^{r_{max}} dE(r,\theta,t)dr \tag{3}$$

where $r = \sqrt{k_x^2 + k_y^2}$, $\theta = \arctan\left(\dfrac{k_y}{k_x}\right)$, and $r_{max} > 0$ is the maximal radial coordinate of the wave number space in the $\theta$ direction.

In order to capture the most prominent relative fluctuations in the EAD, we choose the direction with the largest variation and its perpendicular direction for comparison. These two directions are referred to as the principal direction-pair, and denoted by a 2-tuple $(\Theta_1, \Theta_2)$. In actual data, there may be more than one principal direction-pair. In this case, only one is selected without any preference. We compare the EAD values by:

$$\psi(t) = \ln\frac{\Phi(\Theta_1,t)}{\Phi(\Theta_2,t)} \tag{4}$$

The dimensionless quantity $\psi(t)$ is termed as generalized phase (GP).

In figure 1, both GP and GP amplitude spectra of five different spatiotemporal patterns are calculated. In figure 1 Aa and Ab, a rotating spiral with fixed rotation center and a traveling plane wave are generated by sinusoidal functions in polar and Cartesian coordinate systems respectively. Because of the simple spatiotemporal symmetry of each of these two patterns, only one predominant frequency component arises in their respective GP spectrum (figure 1 Ca and Cb). By our extended definition of the coherent space-time cluster, we say that only one coherent cluster exists in each of these two patterns.

However, it is important to notice that, in general, not every peak in GP spectrum corresponds to a coherent cluster. In neural tissues, origination, extinction, collision and merging of wave fronts are common, and these events frequently change the orientation distribution of wave patterns. Accordingly the number of peaks in the GP spectrum calculated from an actual data set may be more than the number of the coherent clusters arising in wave patterns. To clarify the effects of these dynamics-related events on the estimation of coherent cluster number, we apply the GP amplitude spectrum analysis to different wave patterns with mixture of multiple rotational and translational coherent clusters.

By using a model network of spiking neurons arranged in a 51×51 two-dimensional lattice (see **Appendix B** for model details), wave patterns simulated at different excitation levels are produced for GP spectrum analysis. For the wave patterns simulated at a lower excitation level, two spirals both rotating at around 8Hz stably coexist throughout the whole simulation, during which one or two short-lived spirals or ring waves occasionally arise. Figure 1 Ac illustrates one representative snapshot extracted from the middle of the simulation. 500 consecutive snapshots are extracted for the GP and GP spectrum calculation. The results (see figure 1Bc for GP and figure 1Cc for GP amplitude spectrum) show that the GP spectrum contains three predominant frequency components (the peaks marked with black and gray arrows). Since the orientation of a plane wave before and after rotating 180° is same, the GP oscillation frequency is twice the frequency of the orientation changes in wave patterns. Accordingly, the peak at 17Hz (black arrow) corresponds to the two spirals rotating at around 8Hz, and the other two peaks respectively at around 2Hz and 6Hz (gray arrows) correspond to the changes in orientation distributions caused by events such as collisions, originations and/or extinctions of transient clusters.

*Figure 1 should be inserted around here*

Figure 1Ad illustrates a snapshot extracted from the wave pattern simulated at a higher excitation level. This wave pattern appears chaotic. Spirals constantly break up and merge during propagation and collisions. We extract 500 consecutive snapshots from the simulation for the GP spectrum analysis. The results (see figure 1 Bd for GP and figure 1 Cd for GP amplitude spectrum) show that around five predominant frequency components appear in the GP amplitude spectrum. The movie shows that the simulative wave pattern is highly dynamic. Spirals and rings are frequently originating, colliding, and extinguishing. In most of the 500 consecutive frames, around three to four pairs of coexisting spirals occupy almost the whole sampling region, and are more stable than other clusters. In figure 1 Ad, three pairs of spirals are shown. One pair can be seen clearly on the left half of the snapshot, and the other two pairs (one is slightly above the right middle part of snapshot, and the other is in the right bottom corner) are colliding, and broken wave fronts can be seen clearly. On the left bottom, one transient ring is shown as well.

We tried the GP spectrum analysis on simulated patterns at a variety of different excitation levels, and all results showed similar qualitative accuracy in estimating the number of the coherent clusters by that of the predominant frequency components in GP amplitude spectrum.

Although the aforementioned dynamics-related events seem create more peaks in GP spectrum than corresponding coherent clusters, the number of predominant frequency components is close to the number of coherent clusters. Simulations show that new coherent clusters may be created by collisions of wave fronts, and some of them even become persistent. Moreover, the higher the excitation level the network is, the more wave front collisions create new clusters. Thus, the actual creation of new clusters (short-lived or persistent) partially compensate the overcount caused by these dynamics-related events.

Based on the above observations, it is natural to ask if we can directly use the number of peaks in GP spectrum to approximate the number of the coherent clusters arising in wave patterns. Unfortunately, the answer is still negative, because in addition to the dynamics-related events, in practice, the unavoidable limitations of spatial sampling also cause fluctuations in the EAD, and accordingly lead to extra peaks in GP spectrum. Due to the limited spatial sampling, wave fronts entering and/or leaving the bounded spatial sampling region result in changes in spectral energy of wave patterns. By carefully setting the spatial sampling window, for example, we can make the window sufficiently large relative to the size of single coherent clusters, the effect of limited spatial sampling can be optimally suppressed, and accordingly the amplitude of the extra peaks may be significantly decreased.

*Figure 2 should be inserted around here*

In summary, GP spectrum typically has 4 types of peaks: type-1: peaks corresponding to predominant coherent clusters; type-2: peaks corresponding to dynamics-related events including those corresponding to short-lived coherent clusters; type-3: peaks irrelevant to wave pattern dynamics; type-4: random noise. Numerical calculations on simulative wave patterns with different excitation levels show that the GP spectrum of wave patterns usually contains three different modes: low-frequency mode (Mode I), medium-frequency mode (Mode II), and high-frequency mode (Mode III). Most prominent peaks locate in mode I. In mode II, the GP spectrum shows an exponential decay, which corresponds to the sharp decrease in the occurrence of dynamics relevant events with increasing frequency. In mode III, the GP spectrum is flat, which indicates that only random noises remain in this frequency range. In figure 2A and B, three different modes (separated by two dash-dotted lines) of the GP spectrum of a typical wave pattern (same wave pattern as the one shown in the d column of figure 1) are illustrated. In figure 2B, the GP spectra in mode II and mode III are linearly fitted and shown in solid line and dashed line respectively. The slopes of the two lines are -4.5412 and 0.0908 respectively. In figure 2C and D, the GP spectrum of the spatiotemporal pattern of random noise is linearly fitted, and the slope is $1.5163 \times 10^{-6}$.

From the above discussion, we have seen that type-1 and type-2 peaks are relevant to the dynamics of the system and may correspond to the actual coherent clusters. In order to emphasize their relevance to wave pattern dynamics, we call these types of peaks genuine peaks, and call the other two types of peaks false peaks. Intuitively, short-lived coherent clusters tend to last longer than changes caused by collisions of wave fronts or wave fronts entering or leaving the spatial sampling region, and therefore have more significant effects on changes of spectral energy. Accordingly, dynamics-irrelevant events may be more transient than dynamics relevant events, and have smaller amplitude. Due to the fact that most peaks in mode II are type-2 peaks, by taking half of the maximum amplitude of the peaks in mode II as a threshold, the effects of false peaks may be optimally removed.

The threshold for counting the number of the genuine peaks (NGP) can be defined as:

$$\zeta = \frac{\max_{j \in \{1,2,\cdots,M\}} \{\tilde{\Delta}_j\}}{2} \tag{5}$$

where $\tilde{\Delta}_j$ is the amplitude of the peaks in mode II of the GP spectrum, and $M$ denotes the total number of peaks in mode II. When counting the number of the genuine peaks (NGPs), if $\Delta_i \geq \zeta$, NGPs is increased by one at the $i$-th peak, where $\Delta_i$ is the amplitude of the $i$-th peak in GP spectrum, $i = 1, 2, \cdots, N$, and $N$ designates the total number of peaks in the spectrum.

## III. Quantifying Actual Wave Patterns

By using an EOF based method, we can show that the dimensionality of the neuronal wave patterns consistently decreases during the middle of the episodes of oscillations recorded from the middle layers of a mammalian cortex [7]. Here, we apply the GP spectrum analysis on the neuronal wave patterns recorded from rat cortical slices by voltage-sensitive dye imaging (see [4] and [15] for details of the experimental preparations), and show that the results confirm this observation.

*Figure 3 should be inserted around here*

In figure 3, we calculate the GP (fig 3 first panel) and its power spectrum from a typical imaging trial, and compare the spatiotemporal complexity estimated by different indicators. The wave pattern in this trial can be visualized in a previously published data movie [16]. The initial section of the trial contains less organized irregular waves, with mixed emitting target and spiral waves (0-2500 frames collected at 1629 frames per second). Subsequently, the system becomes less complex with the development of a brief period emitting target waves followed by a period with a well-organized spiral (frames 3800-5200) and a period of plane waves (frames 6200-7000). By the end of the recorded trial, the system becomes complex again with less organized irregular patterns (frames 7500-10000).

The GP power spectrum is calculated by a short-time Fourier transform with a 500-frame sliding window, and plotted in the second panel in time-frequency representation (TFR). The TFR indicates that multiple coherent clusters, including the coexistence of spiral and target waves, occurred in the initial section with irregular activity. In the middle section, target, spiral and plane waves alternatively dominate. The NGP (fig 3, third panel) is close to 1 when the system is dominated by a single wave pattern, whether target, spiral, or plane. The NGP is higher at the beginning and ending sections when multiple coherent clusters originate and die out transiently, and also increases during the reorganizing transition between pattern types (marked with asterisks *).

Visual inspection of the data movie confirms that the NGP and the actual number of coherent clusters agree well with the dynamics of the wave patterns, even when there are only short transitions between the patterns. Compared to KLD and instantaneous effective dimension (IED, [7]), NGP appears to have significantly higher sensitivity in the temporal domain to the changes in wave patterns (fig 3 bottom plots). While the complexity estimated by IED is qualitatively consistent with the NGP, NGP more accurately reflects the number of coherent clusters of these changing wave patterns.

## Discussion and Conclusion

In summary, in the past few decades, various measures have been used to measure the complexity of a system [5]. As a powerful tool for extracting coherent structures, the empirical orthogonal function (EOF) was used to characterize spatiotemporal patterns in neural tissue [6, 7]. Unfortunately however, the orthogonality requirement on the extracted eigenmodes imposes a limit on the physical interpretability. In this paper, we follow the idea of P. Jung et. al. [14], and extend the definition of the coherent space-time cluster to the space-time regions within which wave patterns have identical spatiotemporal symmetry. By inspecting the evolutional features of the orientation distribution [17], we develop a geometric method to estimate the number the coherent clusters. By applying the method on both simulated and actual wave patterns, we show that the NGP gives a qualitatively accurate estimation of the number of coherent clusters.

With notation borrowed from statistical mechanics, P. Jung et. al. [14] defined spatiotemporal entropy of spatiotemporal patterns. With the extended definition of coherence clusters, the spatiotemporal entropy can be defined similarly. It can be shown that with fixed spatial sampling window, the number of coherence clusters arising in wave patterns is proportional to the spatiotemporal entropy of the wave patterns, and using a sliding temporal sampling window, the instantaneous spatiotemporal entropy can be estimated by computing the NGP. Therefore, in comparison with the EOF based methods, the NGP provides a more sensitive and physically relevant measure of the complexity.

## Acknowledgement

This work was supported by NIH Grants R01NS036447 (J.Y.W.) and K02MH04193 (S.J.S). We thank Kay A Robbins for valuable comments on this manuscript.

**Appendix A – Spatiotemporal Symmetry and Coherent Clusters**

Let $f_\Omega(x,y,t)$ be the scalar field of a wave pattern with $(x,y,t) \in \Omega \subset \mathbf{R}^3$. Let the operators of translation symmetry in space and time, rotation symmetry, and translation symmetry in space in radial direction be respectively defined as follows,

$$\kappa_{\vec{\alpha}} : f_\Omega(x,y,t) \mapsto f_\Omega(x+\alpha_x, y+\alpha_y, t),$$

$$\tau_a : f_\Omega(x,y,t) \mapsto f_\Omega(x,y,t+a),$$

$$\rho_\theta : f_\Omega(r,\varphi,t) \mapsto f_\Omega(r, \varphi+\theta, t),$$

$$\kappa_\gamma : f_\Omega(r,\varphi,t) \mapsto f_\Omega(r+\gamma, \varphi, t)$$

where $\vec{\alpha} = <\alpha_x, \alpha_y>$ denotes the displacement vector. Then the spatiotemporal symmetry of wave patterns is defined as follows. The wave pattern $f_\Omega(x,y,t)$ is said to be

1. *translation invariant*, if $\kappa_{\vec{\alpha}} f_\Omega(x,y,t) = \tau_a f_\Omega(x,y,t)$ for some $\vec{\alpha} \in \mathbf{R}^2$, $a \in \mathbf{R}$ and all $(x,y,t) \in \Omega$;

2. *rotation invariant*, if $\rho_\theta f_\Omega(x,y,t) = \tau_a f_\Omega(x,y,t)$ for some $\theta \in \mathbf{R}$, $a \in \mathbf{R}$ and all $(x,y,t) \in \Omega$;

3. *translation invariant in the radial direction*, if $\kappa_\gamma f_\Omega(x,y,t) = \tau_a f_\Omega(x,y,t)$ for some $\gamma \in \mathbf{R}$, $\theta \in \mathbf{R}$ and all $(x,y,t) \in \Omega$.

The region $\Omega \subset \mathbf{R}^3$ in space-time is called a *coherent cluster*, and we use parameters pair $(\vec{\alpha}, a)$, $(\theta, a)$ or $(\gamma, a)$ to characterize the cluster $\Omega$. It is clear that the region $\Omega$ is not necessarily connected. That is, if two disjoint regions $\Omega_1$ and $\Omega_2$ have exactly the same parameters pairs, then they both belong to the same coherent cluster.

**Appendix B – Spiking Neural Network**

Various neuronal network models for simulating wave patterns have been proposed (for review, see [18]). In this paper we use a slightly modified version of the spiking neural network model proposed by W. M. Kistler, R. Seitz, and J. L. van Hemmen [19] for wave pattern generation.

The spiking neural network consists of a 51×51 neuronal lattice, in which each point $(x_i, y_j)$ represents a neuron, and in order to simplify notation, we use $(i, j)$ directly to denote the point $(x_i, y_j)$. The distances between adjacent rows and adjacent columns are both set as 1. The time step in simulations is set as 10ms. The spatiotemporal pattern of the neuronal membrane potentials is given by:

$$V_m(i,j,t) = V_{refr}(i,j,t) + V_{syn}(i,j,t) + V_{stm}(i,j,t) \tag{A1}$$

where

$$V_{refr}(i,j,t) = \sum_{\xi=1}^{t} \eta(t-\xi) S(i,j,\xi) \tag{A2}$$

describes the refractory contribution, and

$$V_{syn}(i,j,t) = \sum_{\substack{m=i-1,i,i+1 \\ n=j-1,j,j+1}} \left( \sum_{\xi=1}^{t} \left( \kappa(t - \Delta_{i,j}(m,n) - \xi) \cdot S(i,j,\xi) \right) \right) \cdot W(m-i, n-j) \tag{A3}$$

describes the synaptic contribution. The stimulation $V_{stm}(i,j,t)$ is set as:

$$V_{stm}(i,j,t) = \begin{cases} 1, & \text{if } 1 \le i \le 25, 24 \le j \le 26, \text{ and } 1 \le t \le 5 \\ 0, & \text{else} \end{cases} \tag{A4}$$

The initial state of the network in simulations at both excitation levels is set as:

$$V_m(i,j,1) = \begin{cases} -0.5, & \text{if } 1 \le i \le 25 \text{ and } 1 \le j \le 23 \\ 0.5, & \text{if } 1 \le i \le 25 \text{ and } 26 \le j \le 51 \\ 0, & \text{else} \end{cases} \tag{A5}$$

In refractory contribution, $\eta(\xi) = -0.1 e^{-\frac{\xi}{\tau_r}} \cdot \Theta(\xi)$ describes the after potential elicited by spikes, and $\Theta(\xi)$ is the Heaviside step function.

In synaptic contribution, $\kappa(\xi) = \frac{\xi}{\tau_s^2} e^{-\frac{\xi}{\tau_s}} \cdot \Theta(\xi)$ describes the form of the postsynaptic potential, and $\Delta_{i,j}(m,n)$ accounts for the time delay of a spike transmitting from neuron $(m,n)$ to neuron $(i,j)$. The synaptic strength of neuron $(m,n)$ coupling to neuron $(i,j)$ is given by:

$$W(m-i, n-j) = \begin{cases} \sum_{k \in \{1,2\}} \alpha_k \cdot e^{-\frac{(x_m - x_i)^2 + (y_n - y_j)^2}{\lambda_k^2}}, & \text{if } |m-i| + |n-j| \ne 0 \\ 0, & \text{else} \end{cases}$$

In both contributions, $S(i,j,t)$ is a spike train of the neuron $(i,j)$, which is described as a sum of the Kronecker delta functions: $S(i,j,t) = \sum_f \delta(t - t_{i,j}^f)$, in which, $t_{i,j}^f$ denotes the firing times of the neuron $(i,j)$, and the index $f = 1, 2, \cdots, N_{i,j}$, numbers the spikes in each spike train, which, in total, contains $N_{i,j}$ spikes. The firing times are defined as: $t_{i,j}^f: V_m(i,j,t_{i,j}^f) = \vartheta$ and $\frac{d}{dt} V_m(i,j,t_{i,j}^f) > 0$, and $\vartheta$ is the threshold for spikes.

For different excitation levels, the following parameters are fixed: $\lambda_1^2 = 15$, $\lambda_2^2 = 100$, $\vartheta = 0.12$, $\tau_s = 4$, $\Delta_{i,j}(m,n)$ is randomly taken as 0, 1, or 2. For higher excitation levels, we take smaller $\tau_r$ and $\alpha_2$, and larger $\alpha_1$. For lower excitation levels, we take larger $\tau_r$ and $\alpha_2$, and smaller $\alpha_1$. For the example shown in fig 1D, we set: $\tau_r = 50$, $\alpha_1 = 1.2$, and $\alpha_2 = -0.25$, and for the example shown in fig 1E: $\tau_r = 2$, $\alpha_1 = 1.4$, and $\alpha_2 = -0.2$.

**Figures and Legends**

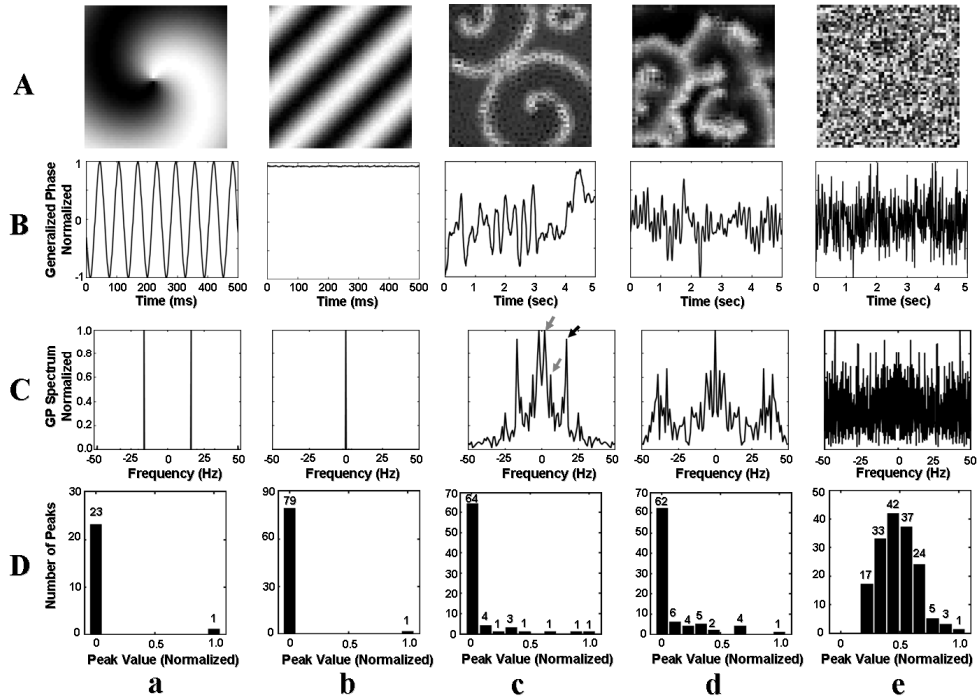

**Figure 1**. The GP amplitude spectrum analysis of a rotating spiral (Aa), a drifting plane wave (Ab), two simulative wave patterns at different excitation levels (Ac and Ad), and a spatiotemporal pattern of uniformly distributed random noises (Ae). In rows B and C, the generalized phase (GP) and GP amplitude spectrum of each spatiotemporal pattern are shown. From both the GP and GP amplitude spectrum, we can see that when the spatiotemporal pattern only contains one single coherent cluster (Aa and Ab), only one single prominent frequency component arises in the GP oscillations (See Ba and Bb for GP traces, Ca and Cb for GP amplitude spectra). (For the ideal drifting plane wave (Ab), in order to avoid $\Phi(\Theta_2,t)=0$, 10% uniformly distributed random noise is added.) When multiple coherent clusters arise (see Ac and Ad), multiple predominant frequency components can be seen in GP oscillations, and comparison between the representative snapshots (Ac and Ad) and the GP spectra (see Cc and Cd) shows that the number of the predominant frequency components is close to the number of the coherent clusters arising in wave patterns (see text for details). Although many peaks exist in the GP amplitude spectrum of the spatiotemporal pattern of random noises (Ce), the distributive features of the peaks is fundamentally different from those of chaotic wave patterns (see Da, Db, Dc, Dd and De for comparison of the distributive features of the peaks in the respective GP amplitude spectra).

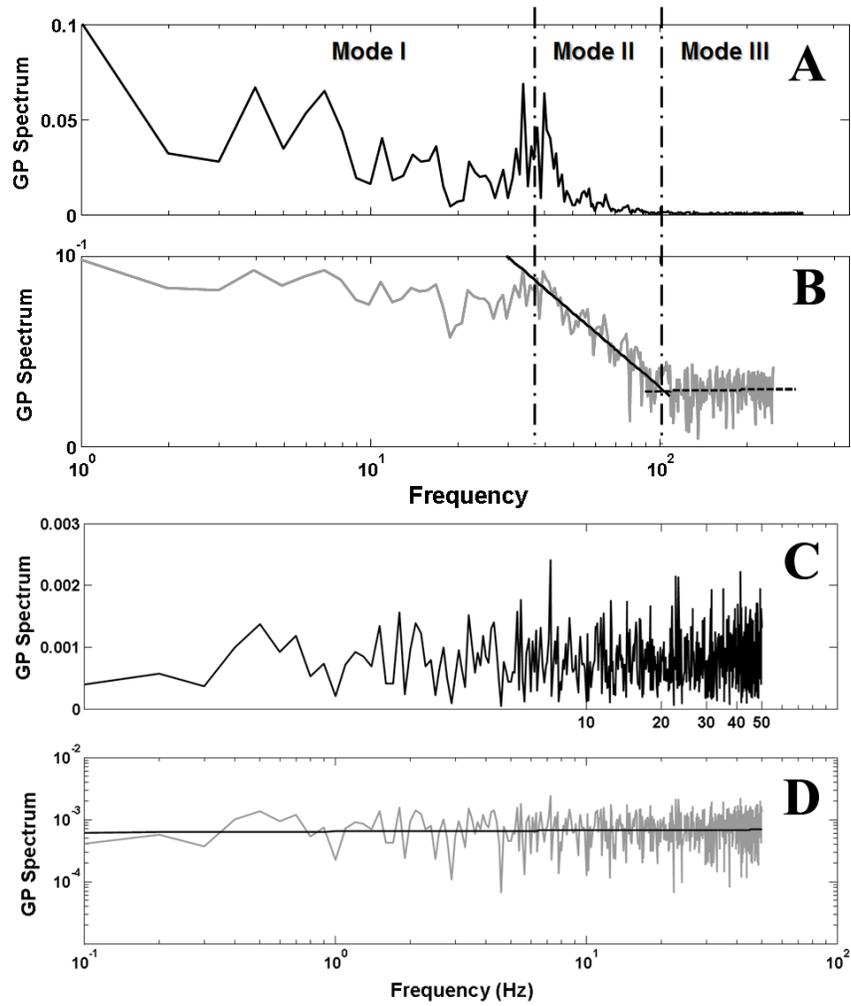

**Figure 2**. Comparing GP spectra of a complex simulated wave pattern with uniformly distributed random noise. A and C are GP spectra shown in Figure 1Cd and Ce respectively.

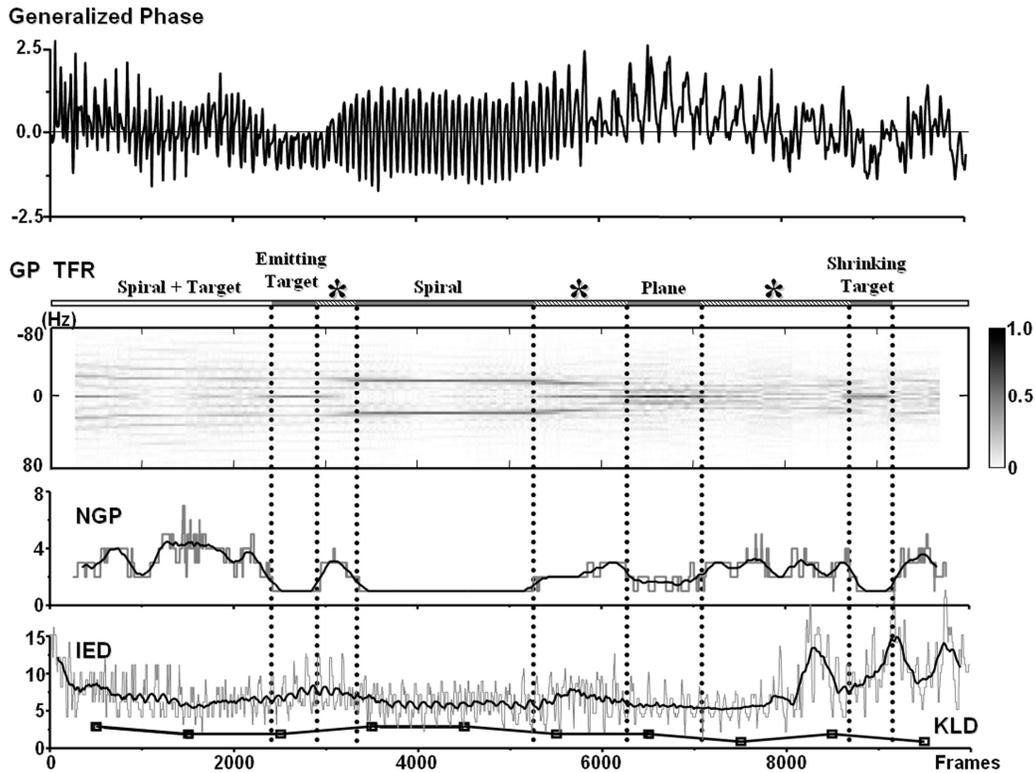

**Figure 3**. Estimating the complexity of cortical wave patterns. First panel: Fluctuation of generalized phase during an imaging trial (~6.2 seconds or 10,000 frames). Middle panel: Time-frequency representation (TFR) of the fluctuating GP. Lower panel: NGP (gray) and its smoothed version (black) obtained by a moving average with a 200-frame window (the parameter $\varepsilon$ is empirically set as 0.4 for counting NGP). Bottom panel: the instantaneous effective dimension (IED, gray) and its smoothed version (black, by moving average with a 200-frame window), and KLD (open squares connected with solid lines) calculated in each of the 10 consecutive segments of the image movie. Each of the segments extracted for KLD calculation consists of 1000 successive frames. Wave patterns labeled over the middle panel (emitting target, spiral …) were identified by visual inspection of the data movie.